\documentclass[sigconf]{acmart}

\usepackage[T1]{fontenc}
\usepackage[utf8]{inputenc}

\DeclareUnicodeCharacter{2212}{\textminus}

\AtBeginDocument{%
  }

 \acmConference[IVA '25]{Preprint version. Accepted at IVA ’25}{Preprint version. Accepted at IVA ’25}{}

\begin{document}

\title{Controlled Yet Natural: A Hybrid BDI-LLM Conversational Agent for Child Helpline Training}

\author{Mohammed Al Owayyed}
\email{M.AlOwayyed@tudelft.nl}
\authornotemark[1]
\orcid{0000-0002-9680-9204}
\affiliation{%
  \institution{Delft University of Technology}
  \city{Delft}
  \country{The Netherlands}}
  \affiliation{
  \institution{King Saud University}
  \city{Riyadh}
  \country{Saudi Arabia}
  }

\author{Adarsh Denga}
\email{A.A.Denga@student.tudelft.nl}
\affiliation{%
  \institution{Delft University of Technology}
  \city{Delft}
  \country{The Netherlands}
  }

  \author{Willem-Paul Brinkman}
\email{W.P.Brinkman@tudelft.nl}
\orcid{0000-0001-8485-7092}
\affiliation{%
  \institution{Delft University of Technology}
  \city{Delft}
  \country{The Netherlands}
  }

 \renewcommand{\shortauthors}{Al Owayyed et al.}

\begin{abstract}

Child helpline training often relies on human-led roleplay, which is both time- and resource-consuming. To address this, rule-based interactive agent simulations have been proposed to provide a structured training experience for new counsellors. However, these agents might suffer from limited language understanding and response variety. To overcome these limitations, we present a hybrid interactive agent that integrates Large Language Models (LLMs) into a rule-based Belief-Desire-Intention (BDI) framework, simulating more realistic virtual child chat conversations. This hybrid solution incorporates LLMs into three components: intent recognition, response generation, and a bypass mechanism. We evaluated the system through two studies: a script-based assessment comparing LLM-generated responses to human-crafted responses, and a within-subject experiment ($N=37$) comparing the LLM-integrated agent with a rule-based version. The first study provided evidence that the three LLM components were non-inferior to human-crafted responses. In the second study, we found credible support for two hypotheses: participants perceived the LLM-integrated agent as more believable and reported more positive attitudes toward it than the rule-based agent. Additionally, although weaker, there was some support for increased engagement (posterior probability = $0.845$, 95\% HDI [–0.149, 0.465]). Our findings demonstrate the potential of integrating LLMs into rule-based systems, offering a promising direction for more flexible but controlled training systems.

\end{abstract}

\ccsdesc[500]{Applied computing~Interactive learning environments}
\ccsdesc[500]{Human-centered computing~Empirical studies in HCI}
\ccsdesc[500]{Computing methodologies~Artificial intelligence}
\keywords{Conversational Agents, Large Language Models (LLMs), Belief-Desire-Intention (BDI), Child Helpline Training, Virtual Training Simulations, Counsellor Training, Social Skills Training}

\maketitle

\section{Introduction}


Child helplines offer a safe space for children in need to seek help. This requires extensive counselling training efforts. Children can contact helplines such as De Kindertelefoon through online chat interfaces or by phone. As of 2023, the Dutch Children's Helpline, De Kindertelefoon, has over 700 volunteer counsellors, with more than 340 new volunteers trained that year—an increase of 20\% compared to 2022 \cite{DKTReport}. To train new counsellors, a common practice is role-playing, where one counsellor acts as a child contacting the helpline while the trainee engages in conversation. However, this method can be time-consuming and resource-intensive. A potential alternative is using interactive social agents to simulate child-helpline interactions, which were helpful in similar training scenarios, such as virtual patients in medical education \cite{guetterman2019medical, yao2022virtual}.

A crucial aspect of a simulation system for learning is being as faithful to the actual situation \cite{saus2010perceived}. In this case, simulating realistic, open-ended chat conversations of a virtual child. However, training counsellors using virtual agents can be sensitive, especially when using open-ended unrestricted interaction \cite{bickmore2022health}. While closed-answer options provide a safer environment, they do not fully replicate real-world interactions where counsellors formulate responses independently to develop their chat conversational skills.

One way to enable open-ended conversations is through rule-based intent recognition, where the system interprets user input and selects appropriate responses. An example of such a system is Lilobot, a BDI-based (Belief-Desire-Intention) training system designed for child helplines \cite{grundmann2025lilobot, al2024cognitive}. The system focuses on bullying scenarios, one of the main topics children discuss with De Kindertelefoon \cite{de-kindertelefoon-report-2023}. Lilobot simulates a virtual child who has been bullied at school and is seeking support from a helpline. The system follows a five-phase model of counselling conversations \cite{counselling-handbook}, which trainees should apply correctly for the interaction to progress. The phases are building rapport, clarifying the child's story, setting the conversation goal, working towards that goal and wrapping up the conversation. If a counsellor fails to follow these phases, Lilobot may leave the conversation. 

However, traditional intent recognition has limitations, as it can only respond within the constraints of its predefined rules and training data \cite{rahman2017programming}. This was also reported for Lilobot. Although positively received for its training potential, participants highlighted certain limitations. Specifically, the system struggled to understand complex sentences outside its knowledge base, indicating a need for improved comprehension. Participants also expressed a need for more response variations, as variety and realism are important for effective learning \cite{atkinson2000learning}. These challenges align with the broader limitations of rule-based conversational agents \cite{rahman2017programming}.

Recent advancements with large language models (LLMs) offer promising solutions to these limitations. LLMs can generate human-like text with contextual relevance and variability, making conversations appear more natural and engaging compared to natural language understanding and generation \cite{freire2024conversational, karanikolas2023large, gurcan2024llm}. However, using pure LLMs in training scenarios presents a challenge; while they enhance realism, they should operate within a controlled learning environment to ensure specific educational objectives are met. Without appropriate constraints, LLMs may generate unintended or extraneous content, which could interfere with the learning process, adding extraneous load to the trainee.

To balance these considerations, we propose integrating LLMs with Lilobot’s existing BDI framework. This paper presents an integration design that preserves the structured learning environment required for counsellor training and evaluates its effects. Our evaluation has two parts: (1) a non-inferiority analysis, conducted through human coders, comparing LLM-generated and BDI system responses in terms of both understanding and response quality, and (2) an experimental study where participants used both LLM-integrated and BDI-only systems in a within-subject design. Based on this evaluation, we formulate and test the following hypotheses:

 \textbf{H1:} Individuals perceive the integrated system to be more believable than the rule-based system.
 
 \textbf{H2:} Individuals perceive the integrated system to be more engaging than the rule-based system.
 
 \textbf{H3:} Individuals have a more positive attitude towards the LLM-integrated system than the rule-based one.

\section{Related Work}

\subsection{Social Agents for Helpline Training}


Researchers explored interactive social agents as tools for training helpline counsellors. Demasi et al. \cite{demasi2020multi} proposed a chatbot simulation to train counsellors in suicide prevention hotlines. This simulation uses annotated transcripts and a multi-task framework to generate responses that mimic various crisis conversation scenarios. Evaluations of the proposed model showed improvements in response diversity and specificity compared to simpler models. For child helpline training specifically, one example is a serious game designed to simulate high-risk interactions in a safe environment (e.g., trafficking and sexual exploitation) \cite{veldhuizen2023use}. This game uses choice-based interactions to train counsellors in handling sensitive situations while developing essential skills. However, it does not support open-ended interactions, which helps avoid language processing and generating limitations.

\subsection{LLMs for Counselling}

LLMs have been used to support counselling in various ways. To simulate a counsellor providing help, Steenstra et al. \cite{steenstra2024virtual} explored LLMs' potential in delivering motivational interviewing for alcohol counselling. They found that LLM-powered virtual agent responses were perceived similarly to human-generated ones, addressing the limitations of rule-based approaches in understanding nuanced therapy conversations. Recently, Heinz et al. \cite{heinz2025randomized} reported a reduction of symptoms in depression, anxiety, feeding and eating disorders after people interacted for four weeks with their therapy chatbot Therabot. This text-based application used a generative LLM fine-tuned on expert-curated mental health dialogues. For training counsellors, Wang et al. \cite{wang2024patient} introduced an LLM-driven virtual patient to train trainees in formulating cognitive models as part of cognitive behavioural therapy. The authors defined several cognitive models as scenarios and instructed the LLM to simulate patient behaviour using them. This also allows trainees to compare the identified cognitive model with the actual one. When evaluating the system, the simulated interactions resulted in greater perceived knowledge and confidence compared to traditional methods. Other examples of LLM-integrated virtual patients include applications in history taking \cite{holderried2024generative}, and medical diagnosis \cite{scherr2023chatgpt, vaughn2024enhancing}. LLMs have also been used to support counsellors in their work or training by generating reflections \cite{shen2020counseling}, providing feedback on trainees' responses \cite{brugge2024large, chaszczewicz-etal-2024-multi}, and offering suggestions to counsellors \cite{lai2023supporting}.

\subsection{Integrations of LLMs and BDI}

The integration of LLMs with rule-based systems has been explored to enhance simulations' reliability, realism, and explainability. Pico et al. \cite{pico2024exploring} examined integrating LLMs with a BDI model to improve emotion recognition in intelligent agents. In their approach, an LLM is prompted with a dialogue and tasked to pick up an emotion, which is then converted into beliefs used by the BDI model. When evaluated, the LLM demonstrated promising capabilities in emotion recognition. Frering et al. \cite{frering2025integrating} integrated LLMs and BDI in a human-robot interaction setting to improve explainability. The authors used an LLM to interpret user input and translate it into commands the BDI agent could process. Additionally, the LLM generated responses explaining the robot's behaviour based on the BDI agent’s state.

\section{Integrating LLMs with Lilobot}

We integrated LLMs into Lilobot's rule-based framework to enhance understanding and generation capabilities while maintaining control over scenario structure. We examined integration possibilities within the existing system, explored LLM models and prompt strategies, and then selected a practical implementation approach.

\subsection{Integration Possibilities}

Lilobot follows a BDI architecture \cite{grundmann2025lilobot}, which defines internal states for the virtual child, making its decision-making controllable and interpretable. The system maps trainee input to predefined intents—for example, when a trainee writes to the virtual child, “How does that make you feel?”, the input is recognised as the intent “request\_unknown\_feeling.” These intents are mapped to child beliefs within the BDI framework, which are tied to the child's desires and intentions—for instance, the intent will increase the belief that “they feel the trainee can be trusted.” Based on the current BDI state and the recognised trainee intent, the system picks a response from its knowledge base. To introduce variation, each response has four alternatives, one of which is selected at random; in this case, the child might reply, “It makes me sad... I really don't know what to do.” If the trainee’s intent is not recognised, the system retrieves a default response based on Lilobot’s currently active desire (i.e., goal). For example, when the desire “they want the trainee to call their school” is active, a default response could be “I want you to call my school.” While this structure ensures consistency, it also limits the dialogue’s flexibility and adaptability to diverse responses.

To identify where we can integrate LLMs, we followed the general architecture of agent-based training for social skills (ARTES) \cite{alowayyed2024} and Lilobot's limitations. ARTES outlines three possible integration points in the simulation: Natural Language Understanding (NLU), Natural Language Generation (NLG), and Lilobot's thinking. NLU and NLG have been shown to perform well with LLMs \cite{chang2024survey}, allowing for a better understanding of trainee input as well as increased variability and naturalness in responses. As for the thinking component, integrating an LLM would enable Lilobot to bypass unexpected cases not covered in the list of trainees' intents. Therefore, we refer to it as the Bypass component in this paper.

\subsection{Components}

\begin{figure}[ht]
\centering
    \includegraphics[width=.95\linewidth]{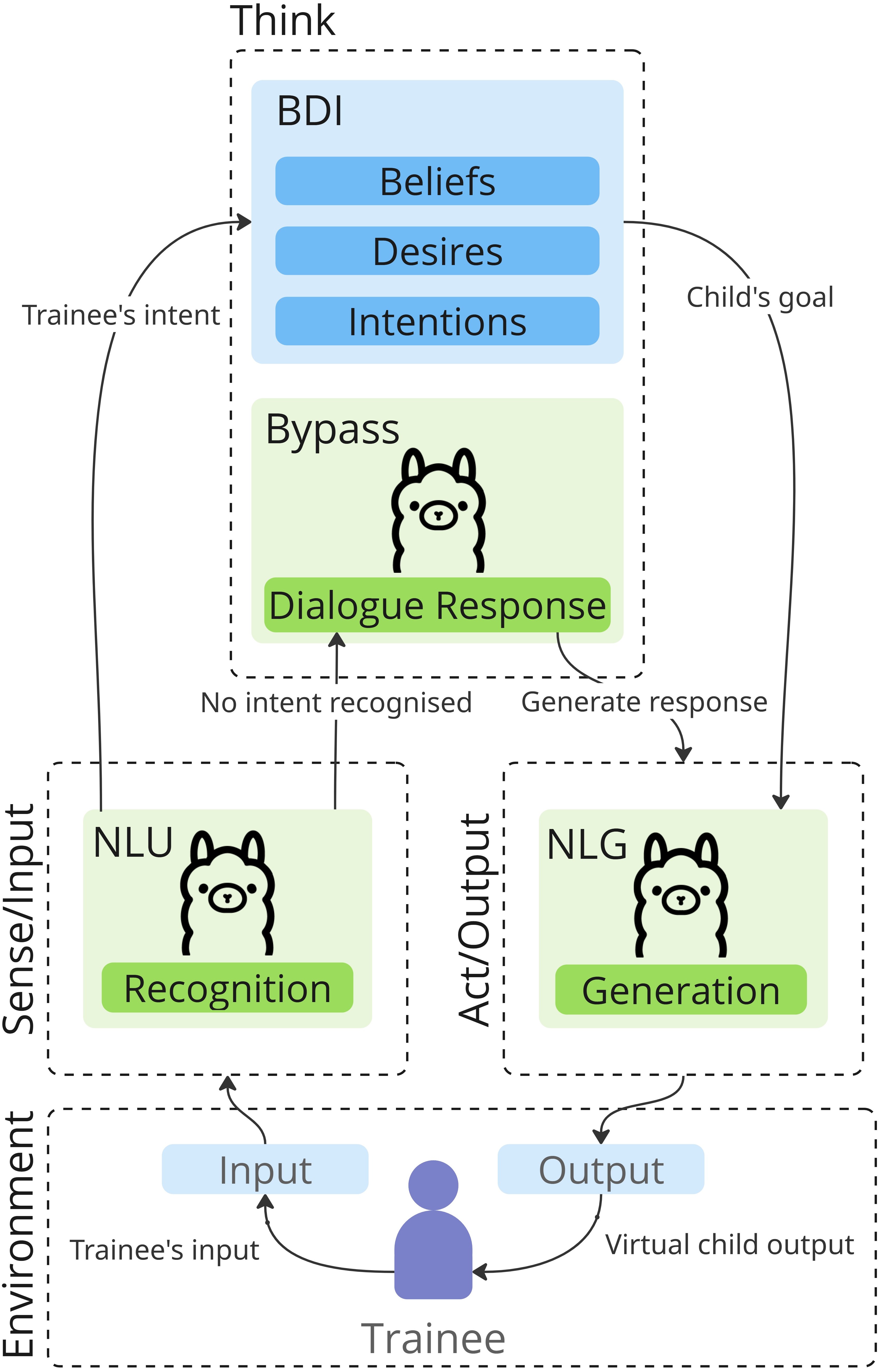}
    \caption{Architecture of the LLM-integrated BDI-based conversational agent based on the ARTES architecture \cite{alowayyed2024}. The green components indicate where the LLM is integrated.}
    \label{fig:Arch}
\end{figure}

Figure \ref{fig:Arch} shows the integrated system's architecture. The overall flow begins with the trainee sending an input, which the NLU component matches to a trainee intent using over 2,000 examples from a dataset used by the rule-based system. If a match is found, the intent is sent to the BDI system to update the cognitive state of the virtual child, resulting in the identification of a response with four associated example responses. These are passed on to the NLG component along with the original input message from the trainee. The NLG component, in turn, generates a contextually appropriate response for the trainee. If the NLU finds no matching intent, the Bypass component generates a relevant response based on the trainee's input, which is then sent back to the trainee.

\subsection{Prompt Creation}

We explored several prompts for the three components: NLU, NLG, and Bypass. We followed prompt engineering techniques \cite{arvidsson2023prompt, udemyPromptEngineering2024}. This led to specific prompt designs, one for each component.

\textbf{NLU:} To classify trainee input into a BDI-recognisable intent, we designed a prompt in which the LLM’s task is to select the best-matching intent for the input or return \textit{``unknown''} if no match is found. When a trainee input is received, it is embedded and compared—using L2 distance in embedding space—to over 2,000 annotated examples (e.g., \textit{``why are you being bullied?''} mapped to the intent \textit{bullying\_why}), which are stored in a vector database. The closest examples to the trainee's input are then included in the prompt, along with their corresponding intents and the trainee's input itself, and sent to the LLM. This method reduced the prompt size while preserving contextual relevance. In our initial attempts, we embedded all examples in the prompt, which resulted in slower response times.

\textbf{NLG:} When a trainee's input is matched to an intent, it is sent to the BDI in the Think component, along with the child's state, and then passed to the NLG sub-component to generate a response. The NLG prompt includes the trainee’s input, a defined child persona (e.g., \textit{``You must play the character of Lilo, a 9 year old child being bullied at school.''}), the child’s current goal as represented in the BDI model, and four example child responses linked to the identified trainee intent. The prompt instructs the LLM model to generate a reply similar in tone and structure to these examples. The resulting response is then sent to the trainee as the virtual child’s reply.

\textbf{Bypass:} When no matching trainee intent is found, the Think component triggers the Bypass sub-component. The prompt structure is similar to that of the NLG prompt, with the key difference being the absence of example responses. In this case, the LLM is instructed to generate a contextually appropriate response from a child's perspective without relying on predefined examples. The generated response is then sent to the trainee.


\section{Study 1: Script Evaluation}

To evaluate the effectiveness of the prompts, we assessed whether the outcomes generated by the LLM were non-inferior to those produced by human-crafted responses. These human-crafted responses, drawn from Lilobot’s rule-based model, had been reviewed by domain experts \cite{grundmann2025lilobot}. Inspired by Steenstra et al. \cite{steenstra2024virtual}, we employed a within-subjects design in which we asked four human coders to rate or label the outcomes of both the LLM prompts and the human-crafted responses from the rule-based system. The coders included an undergraduate medical student, a computer science graduate, and two master's students in computer science. For NLU, we focused on categorisation accuracy. We created 12 trainee input examples, where the LLM prompt, the rule-based system, and the coders independently matched each utterance to a trainee intent, or labelled it as "unknown" if it did not match any trainee intent. For NLG and bypass, we evaluated whether the language used in the LLM-generated responses was appropriate to the context, the virtual child's state, and the trainee’s input. We generated eight NLG responses per system: eight human-crafted responses from the rule-based system and eight LLM-generated counterparts. Additionally, we generated eight LLM bypass responses that could not be matched to intents in the rule-based system. Participants were then presented with all 24 responses in randomised order, along with the conversation history for context, and were asked to rate them. The source of each response was concealed from the participants. All data, prompts, pipelines for the prompts, and analysis code are available online \footnote{https://data.4tu.nl/datasets/5c5b69b7-f727-42d0-983f-07ab910b8460}. 

\subsection{Measures}

\textbf{NLU:} Participants were given a list of 38 trainee intents with brief explanations and 12 trainee inputs. Their task was to match each input to one or more intents. In the ground truth, 10 inputs were each mapped to a different intent, and the remaining two were mapped to two intents each, resulting in 14 intents to identify.

\textbf{NLG and Bypass:} Participants were asked to rate each response on a Likert scale (1 = Strongly Disagree to 7 = Strongly Agree) based on five statements adapted from Steenstra et al. \cite{steenstra2024virtual} (Table \ref{table:analysis-NLG-Bypass-criteria}).

\begin{table}[ht]
\centering
\caption{Rating statements for the Bypass and NLG evaluation tasks, adapted from \cite{steenstra2024virtual}.}
\label{table:analysis-NLG-Bypass-criteria}
\renewcommand{\arraystretch}{1.2}
\begin{tabular}{p{0.95\columnwidth}} 
\hline
\textbf{Statements} \\
\hline
C1: This response is in coherent English \\
C2: This response is coherent in this bullying context \\
C3: This response directly addresses and replies to the counsellor’s previous message \\
C4: This response makes sense \\
C5: This response makes sense in this context of bullying \\
\hline
\end{tabular}
\end{table}

\subsection{Data Analysis}

We conducted a Bayesian analysis to perform a non-inferiority test. Specifically, we fitted a Bayesian multilevel regression model, using non-informative priors. For NLU, we used the model to estimate whether the intent categorisation result for each example matched the ground truth using the rule-based model or the LLM. For the NLG and Bypass components, the model accounted for variability across individual coders when rating LLM- and human-crafted utterances. In both cases, we defined a non-inferiority threshold just below the null value. We then assessed the proportion of the posterior distribution of the parameter estimate that fell above this threshold, which would indicate practical non-inferiority \cite{kruschke2018rejecting}. The threshold was set based on half of Cohen’s \cite{muller1989statistical} threshold for a small effect size ($-0.1 \times \mathrm{sd}(\text{score})$) to define negligible or better effects \cite{kruschke2018rejecting}. Additionally, inter-rater reliability was assessed using the Intraclass Correlation Coefficient (ICC), Fleiss' and Cohen's Kappa to estimate agreement between coders, the LLM, the rule-based system, and ground truth.

\subsection{Results and Discussion}

In the NLU analysis, the LLM correctly recognised all 14 trainee intents for the 12 examples, while the rule-based system recognised 12 intents, and the median of the four human coders recognised 13. To check for overall reliability, Fleiss’ Kappa among human coders alone was 0.81, and between the coders and the LLM was 0.82, which Landis and Koch \cite{landis1977measurement} classify as “almost perfect” agreement. Agreement between coders and the rule-based model was slightly lower but still substantial ($\kappa = 0.78$), according to the same classification scale. Cohen’s Kappa indicated perfect agreement between the LLM and ground truth ($\kappa = 1.00$), and almost perfect agreement between the median coder and ground truth ($\kappa = 0.92$), as well as between the rule-based model and ground truth ($\kappa = 0.85$) \cite{landis1977measurement}. The non-inferiority results indicated some support for the non-inferiority of the LLM, with the probability of 0.87 being non-inferior.

For NLG and Bypass, coders' ratings agreement was generally low (ICC $<$ 0.5). Median ratings clustered near the top of the scale (coder 1 = 6.5, coder 2 = 7, coder 3 = 7, coder 4 = 5.4), potentially indicating ceiling effects and limited variance. LLM texts received a mean rating of 6.30 (SD = 0.89), while human-crafted texts had a mean of 6.28 (SD = 0.78). The results showed a posterior probability of 0.70, indicating some support for the non-inferiority of LLM-generated content relative to human-crafted text.

\section{Study 2: Experiment}

After evaluating the LLM-generated script quality, we conducted a within-subjects experiment to test our hypotheses. Participants role-played as counsellors tasked with applying the five-phase model with the virtual child. There are two conditions: (1) a rule-based agent and (2) an LLM-integrated agent. Each participant experienced both systems in a randomised order to control for order effects, with the entire experiment lasting approximately 50 minutes. This experiment was approved by the university human research ethics committee (HREC reference number: 4768) and preregistered with the Open Science Framework\footnote{https://osf.io/eqxwz}. The data, analysis code in R (version 4.3.1), and a Docker container for reproducibility are publicly available via the 4TU research data repository \footnote{https://data.4tu.nl/datasets/5c5b69b7-f727-42d0-983f-07ab910b8460}.

\begin{figure*}[h]
    \centering
    \includegraphics[width=1\linewidth]{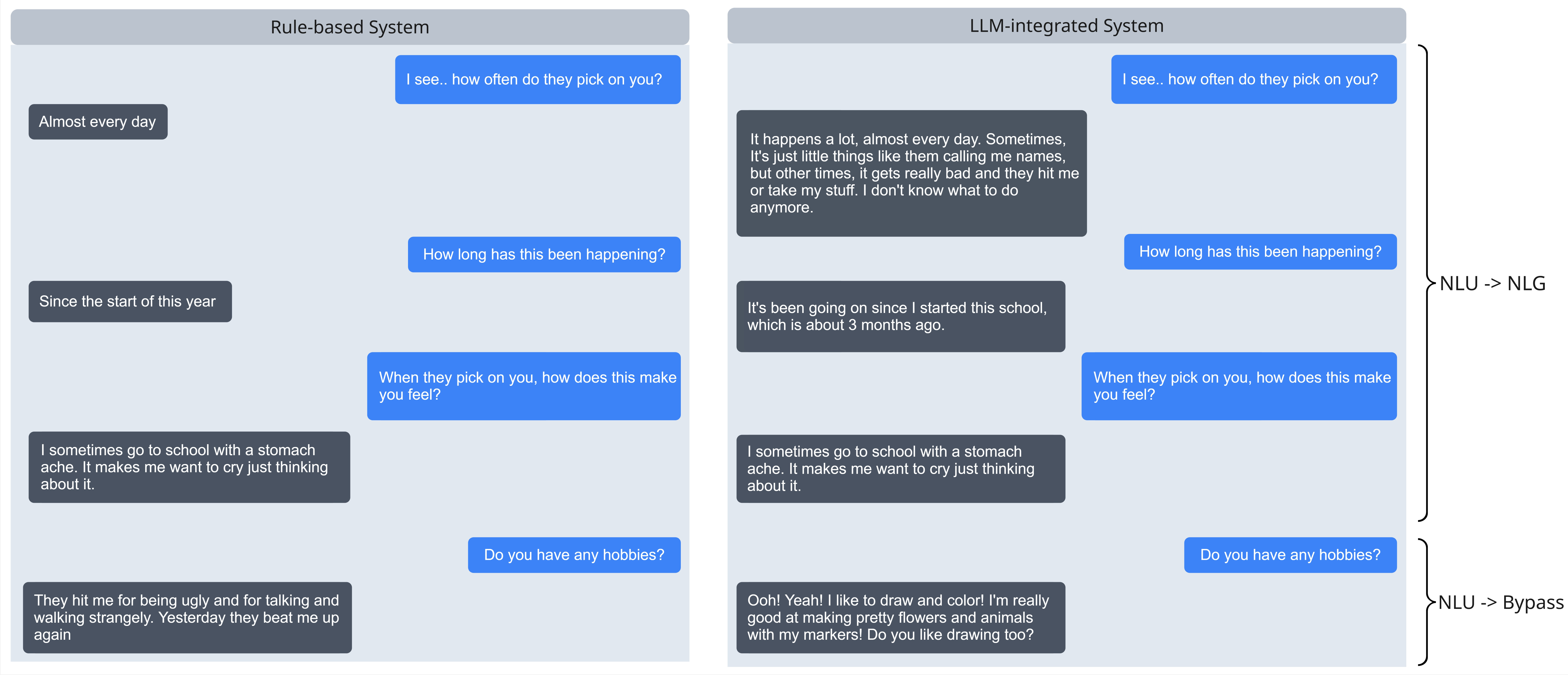}
\caption{Example conversation for each condition. The rule-based system (left) and the LLM-integrated system (right) respond differently to the same counsellor input. The last input does not exist in the trainee intents list, triggering the Bypass component in the LLM-integrated system. Meanwhile, the rule-based system responds with a message related to the child's active desire.}
    \label{fig:screenshots}
\end{figure*}

\subsection{Participants}

We recruited participants through the online platform Prolific. To match the target group for this training system, e.g., new helpline volunteers at organisations like De Kindertelefoon, we determined eligibility as adults with at least a Bachelor's degree. We also required fluency in English, an approval rating of \( \geq 95\text{\%} \), and have completed at least five studies in the past. A total of 55 participants were invited; 18 were excluded in total due to incomplete participation (n = 13), failing comprehension or attention checks (n = 2), not providing consent (n = 1), or not meeting screening criteria (n = 2). This resulted in a final sample of 37 participants. The study was conducted between January and February 2025. Participants received compensation per Prolific’s minimum payment policies.

Of the 37 participants, 24 held a Bachelor’s degree or equivalent (65\%), 10 held a Master’s degree (27\%), and 3 held a PhD (8\%). Fourteen participants identified as male (38\%) and 23 as female (62\%). Age distribution was as follows: 18–24 (n = 10), 25–34 (n = 12), 35–44 (n = 9), 45–54 (n = 5), and 55–64 (n = 1). Participants represented diverse nationalities, including the United Kingdom, Portugal, the United States, South Africa, Nigeria, and India.

\subsection{Measures}

We tested primary and secondary measures to evaluate the rule-based and integrated systems. The primary measures were used to test the three hypotheses, while the secondary measures provided additional insight into participants’ experiences.

\subsubsection{Primary Measures}

We have three measures corresponding to the hypotheses: Believability, Engagement, and Attitude. All questionnaire items were drawn from the Artificial Social Agent  Questionnaire (ASAQ) \cite{fitrianie2022artificial, fitrianie2025artificial}, using a 7-point scale from –3 (Disagree) to +3 (Agree), with 0 as Neither agree nor disagree. We replaced references to “the agent” with “the virtual child” in all questionnaire items.

\textbf{Believability:} Measured using two constructs from the ASA questionnaire, Human-Like Behaviour (HLB) and Natural Behaviour (NB). They capture how believable the agent is to participants.

\textbf{Engagement:} Measured using the engagement construct, assessing participants’ involvement during the interaction.

\textbf{Attitude:} Measured using the attitude construct, reflecting participants’ feelings toward the virtual child.

\subsubsection{Secondary Measures}

To gain further insight, we included the following secondary measures:

\textbf{Overall Experience:} Assessed using the 24-item short version of the ASA questionnaire, capturing broader constructs evaluation, expressed in short ASAQ score.

\textbf{Preference:} At the end of the experiment, participants answered a preference question: “Which virtual child did you prefer interacting with?”, requiring a selection between "The first child" or "The second child". 

\textbf{Experience:} Qualitative feedback was collected through an open-ended question after each condition (“How did your interaction with the virtual child go?”).

\subsection{System}

Both conditions (rule-based and LLM-integrated) were presented through an identical web interface hosted on TU Delft servers. The layout displayed the conversation on the right and a summary of the five-phase model on the left. Each participant interacted with two of the twelve bullying-related scenarios, with each virtual child having a different, randomly assigned name. Scenarios were randomised across conditions to avoid systematic scenario bias and were assigned such that no participant encountered the same scenario twice. As described above, the systems differed only in how they processed and responded to the participants' input. The rule-based agent used Rasa for NLU and predetermined responses, while the LLM-integrated version used Llama 3.2, run via Ollama, for NLU, NLG, and bypass. To ensure a fair comparison, we modified the rule-based system to match the response time of the LLM-integrated system (a random time between 15-25 seconds). Figure \ref{fig:screenshots} shows a conversational example from the rule-based system (left) and the LLM-integrated one (right). The code is publicly available online\footnote{https://github.com/adarshdenga/llm-integration-childhelpline}.

\subsection{Procedure}

Participants were redirected from Prolific to Qualtrics, where they completed an informed consent form. They then watched a short training video introducing the five-phase model and providing context on counselling practices. Afterwards, they interacted with the two systems (rule-based and LLM-integrated), with the order chosen at random. Participants were first directed to one of the systems and instructed to interact with it for 15 minutes. Within this time, the scenario restarted if the virtual child exited or if the participant ended the interaction (e.g., by saying “bye”). After 15 minutes, participants were redirected to Qualtrics to complete questionnaires on believability, engagement, attitude, overall experience, and an open-ended reflection question based on the virtual child they had just interacted with. They were then redirected to the second system, following the same structure: 15 minutes of interaction, followed by the same questionnaires based on the second virtual child. At the end, participants chose their preferred virtual child.

\subsection{Data analysis}

We analysed data from the 37 participants who were included in the study. We reversed items where required for all questionnaire-based measures, according to the ASA questionnaire guidelines, and computed average scores per construct. To test the hypotheses, we conducted Bayesian\footnote{
We reported the Bayesian analysis results in the main text to get more insight and to be consistent with Study 1. In the preregistration, we indicated using a frequentist test by specifying a cutoff value of \( p < 0.05 \). For transparency, the frequentist results for each measure are as follows:

\begin{itemize}
    \item \textbf{Human-like Behaviour (H1):} \( t(37) = 2.10, \, p = 0.04^{*} \)
    \item \textbf{Natural Behaviour (H1):} \( t(37) = 1.38, \, p = 0.18 \)
    \item \textbf{Engagement (H2):} \( t(37) = 1.07, \, p = 0.29 \)
    \item \textbf{Attitude (H3):} \( t(37) = 2.46, \, p = 0.02^{*} \)
    \item \textbf{Overall Experience:} \( t(37) = 2.57, \, p = 0.01^{*} \)
    \item \textbf{LLM Preference:} $\hat{p} = 0.70$, 95\% CI [0.53, 0.84], $p = .02$\textsuperscript{*}
    
\end{itemize}

 } paired-sample $t$-test using non-informative priors. We calculated the posterior probability that the difference between the averages (LLM-integrated – rule-based) is greater than zero and interpreted it based on the guidelines by Chechile \cite{chechile2020bayesian} and Andraszewicz et al. \cite{andraszewicz2015introduction}. We also checked whether zero is included in the credible intervals expressed by the Highest Density Interval (HDI). For the preference question, we used a Bayesian binomial test. We performed a thematic analysis \cite{Braun01012006} of the qualitative responses to identify recurring patterns and insights across the open-ended questions. To ensure reliability, two coders with a background in computer science independently coded the responses using a predefined coding scheme. The inter-coder reliability was fair (Cohen's $\kappa = 0.32$) \cite{landis1977measurement}. A third coder then reviewed the disagreements to determine the final coding.

\subsection{Results}

\begin{figure*}[h]
    \centering
    \includegraphics[width=0.9\linewidth]{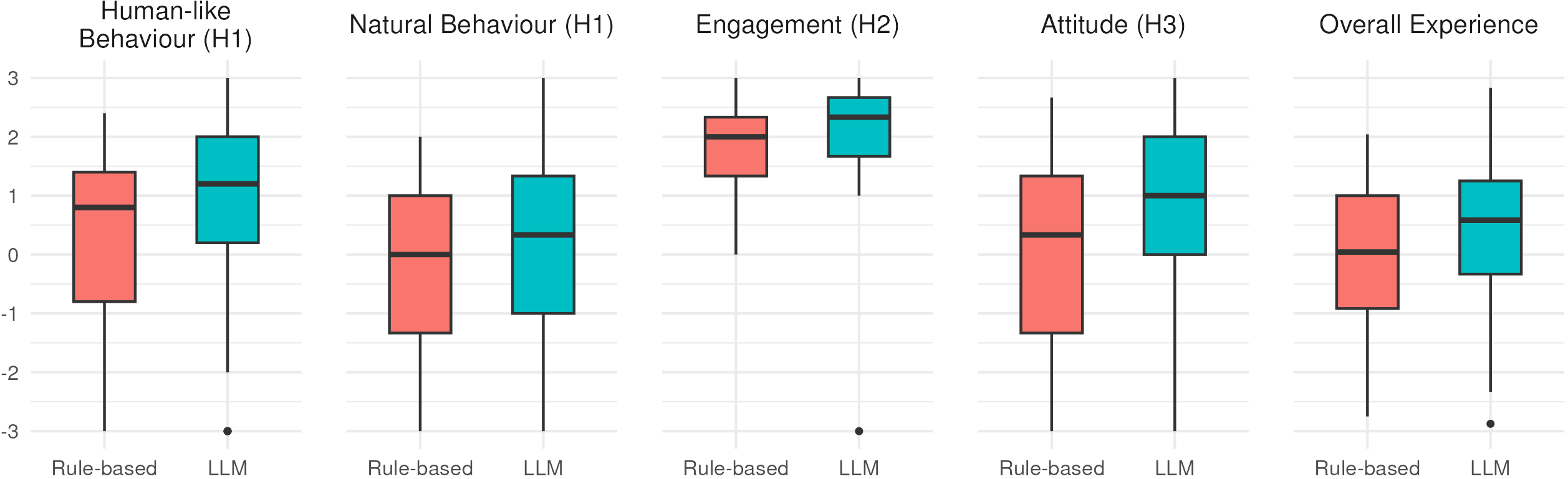}
    \caption{Comparison of participant ratings across five constructs between the rule-based and LLM-integrated agents.}
    \label{fig:boxplots}
\end{figure*}

\subsubsection{Primary Measures} Figure \ref{fig:boxplots} shows an overall comparison across all measures. The overall trend seems to favour the LLM-integrated system.

\begin{figure}
    \centering
    \includegraphics[width=0.95\linewidth]{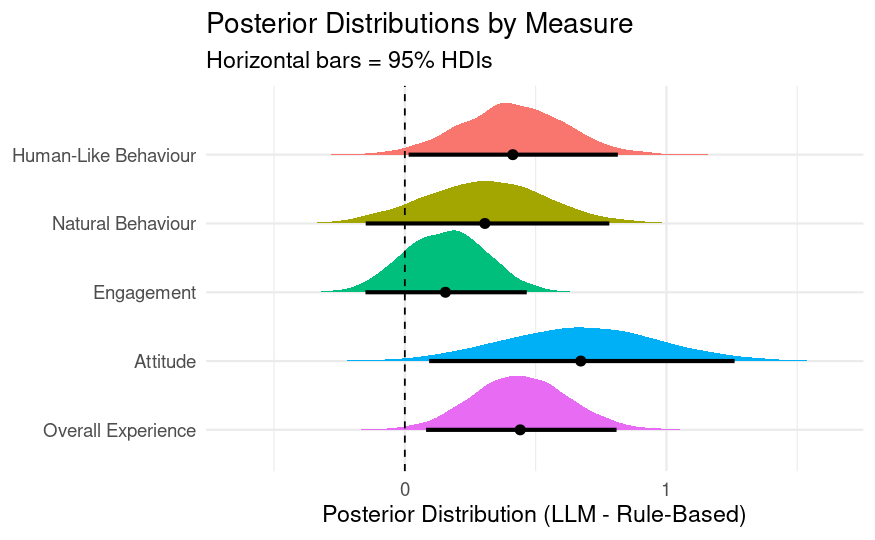}
    \caption{Posterior distributions of effect sizes comparing the LLM-integrated to the rule-based system across outcome measures. }
    \label{fig:posterior-dist}
\end{figure}

\textbf{Believability (H1)}
Human-like behaviour and natural behaviour were rated higher for the LLM-integrated system (HLB: M = 0.79, SD = 1.62; NB: M = 0.14, SD = 1.55) than the rule-based system (HLB: M = 0.35, SD = 1.57; NB: M = −0.19, SD = 1.57). The posterior probability for the human-like behaviour was 0.975, a good bet in favour of the LLM system, with a 95\% HDI of [0.002, 0.816] which excludes zero. This further supports the difference as the HDI range is above zero (Figure \ref{fig:posterior-dist}). For the natural behaviour, the posterior probability was 0.905, which is considered a promising but risky bet for the LLM-integrated system. This is also shown in the 95\% HDI [-0.161, 0.771], which overlaps with zero. Given these two results, there is credible support for H1.

\textbf{Engagement (H2)}
Engagement ratings were slightly higher for the LLM-integrated system (M = 2.01, SD = 1.05) than the rule-based system (M = 1.84, SD = 0.77). The posterior probability of 0.845 constitutes only a casual bet, with 95\% HDI [-0.149, 0.465] included zero (Figure \ref{fig:posterior-dist}); thus, there is some support for H2.

\textbf{Attitude (H3)}
Participants reported a more favourable attitude toward the LLM system (M = 0.86, SD = 1.52) than the rule-based system (M = 0.13, SD = 1.76), with a posterior probability of 0.988—a good bet for the LLM system. The 95\% HDI [0.083, 1.266] is above zero (Figure \ref{fig:posterior-dist}). This indicates credible support for H3.

\subsubsection{Secondary Measures}

Similar to the primary measure, the overall trend seems to favour the LLM-integrated system.

\begin{figure}
    \centering
    \includegraphics[width=1\linewidth]{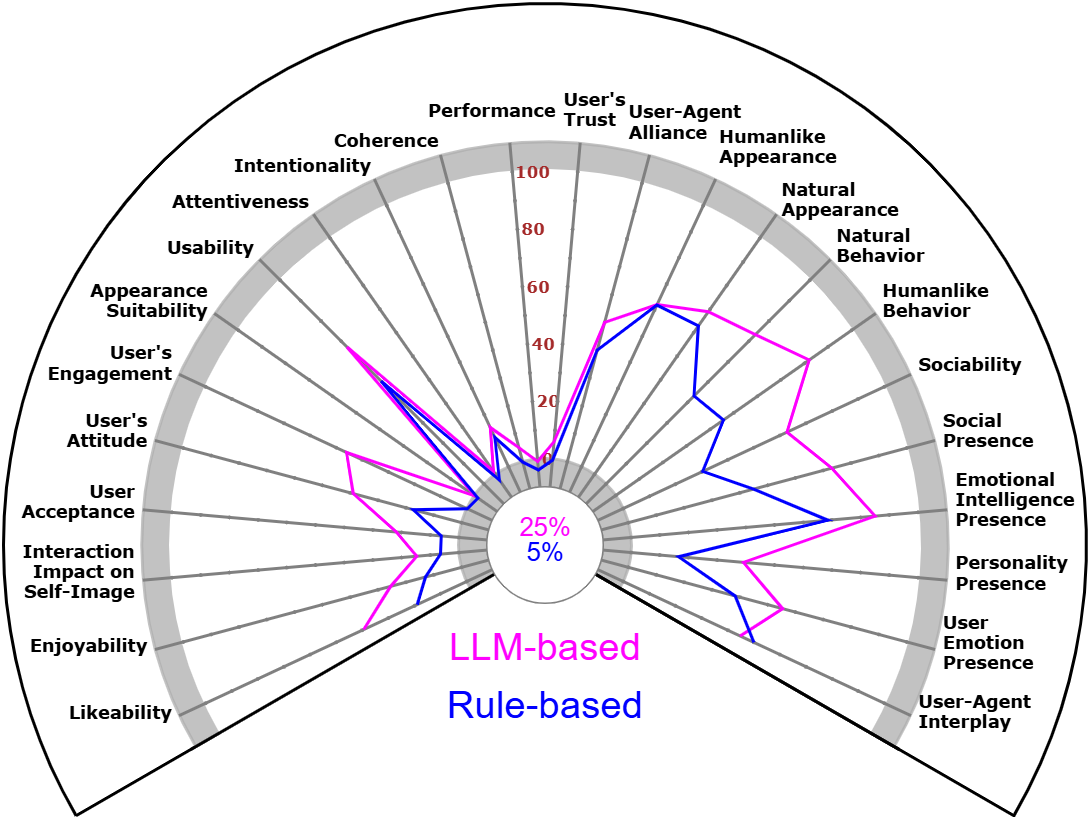}
    \caption{Percentile chart comparing the LLM-integrated and rule-based agents against ASAQ Representative Set 2024 \cite{fitrianie2025artificial}. The grey area represents the range of scores higher or lower than the 29 agents in the dataset. The visualisation was created using ASAQ's visualisation tool \cite{fitrianie2025artificial}.}
    \label{fig:asa_comparison}
\end{figure}

\textbf{Overall Experience}
The overall experience rating is also higher for the LLM-integrated system (M = 0.46, SD = 1.25, Short ASAQ score = 11) than the rule-based system (M = −0.021, SD = 1.28, Short ASAQ score = -1). The posterior probability was 0.991—a strong bet and irresponsible to avoid, with the 95\% HDI [0.075, 0.807] higher than zero (Figure \ref{fig:posterior-dist}). We also used the ASAQ Representative Set 2024, consisting of 29 artificial social agents, for broader cross-study benchmarking, as suggested by the ASAQ's authors \cite{fitrianie2025artificial}. This resulted in Figure \ref{fig:asa_comparison}, where the LLM-integrated agent outperformed 25\% of the set, while the rule-based agent exceeded only 5\%.

\textbf{Preference}
The preference analysis indicates a preference for the LLM system, with 26 participants favouring it. The posterior probability that this preference is above chance is 0.99, which is a strong bet that the LLM-integrated agent is preferred.

\textbf{Qualitative Feedback} Table \ref{tab:thematic_Analysis} presents nine identified themes and their frequency in each condition. The themes relate either to the child's replies (the first five themes) or to the interaction experience (the last four). Overall, positive themes were more prevalent in the LLM condition, such as perceived conversational depth and human-like responses (e.g., \textit{``\ldots{} I think it acted of of a human being. It was quite impressive''}). In contrast, negative responses were more frequent in the rule-based condition, including unnatural and slow responses (e.g., \textit{``It was boring, non-responsive, had delayed responses, did not reveal any realistic feelings, etc.''}). One theme observed in both conditions—abrupt endings—may be partly attributed to the controlled scenario design, in which the virtual child could exit the conversation if the trainee deviated from the five-phase model (e.g., failing to show empathy or directly suggesting an action). 

\begin{table}[h]
\centering
\caption{Themes identified from participants’ responses and the frequency of their occurrence in each condition.}
\label{tab:thematic_Analysis}
\begin{tabular}{lcc}
\toprule
\textbf{Theme} & \textbf{LLM-integrated} & \textbf{Rule-based} \\
\midrule
Abrupt Ending         & 3  & 7  \\
Unnatural Responses   & 4  & 7  \\
Depth of Conversation & 4  & 0  \\
Emotional Engagement  & 6  & 1  \\
Human-Like Responses  & 4  & 2  \\
Positive Experience   & 9  & 8  \\
Scripted Responses    & 0  & 1  \\
Slow Responses        & 6  & 10 \\
Boring Task           & 1  & 1  \\
\bottomrule
\end{tabular}
\end{table}

\subsection{Discussion}

In this study, we evaluated whether integrating LLMs into a rule-based system could enhance believability (H1), engagement (H2), and positive trainee attitudes (H3) toward a virtual child used for helpline counsellor training.

We found credible support for believability (H1), particularly human-like behaviour. Furthermore, we hypothesised greater engagement for the LLM-integrated system than the rule-based one (H2). The results provided some support for this. Although scores were slightly higher, the posterior distribution indicated only a casual probability in favour of the LLM system. This may be attributed to a ceiling effect, as both systems scored highly on engagement (Figure \ref{fig:boxplots}). Such high engagement may result from the inherently engaging nature of the task, that is, providing support to a child in need. Additionally, participant comments clearly indicated task-related engagement (e.g., for the rule-based system: \textit{``\ldots{}we worked together to talk about their feelings and challenges at school. It was a productive conversation\ldots{}''}, and for the LLM-integrated system: \textit{``\ldots{}The virtual child was engaging, and we had a good conversation! I felt like we made a connection, \ldots{}''}). 

Regarding attitude, we found credible support that people rated the LLM-integrated agent more positively than the rule-based agent (H3). Participants also directly compared the two systems in their comments, with three participants explicitly indicating that the LLM-integrated interaction was better (e.g., \textit{``it went better than the first one [i.e., the rule-based]\ldots''} and \textit{``Definitely better than during the 1st chat [i.e., the rule-based]\ldots''}).

Looking beyond the results obtained from hypothesis testing, we observed that participants had a more favourable overall experience with the LLM-integrated agent. Figure \ref{fig:asa_comparison} shows that the LLM-integrated agent scored above the 75th percentile on constructs related to social presence and emotional intelligence. The agent’s emotional presence was also supported by our thematic analysis (e.g., for the LLM-integrated system: \textit{``\ldots{}the virtual child was attentive and behaved more like a human and there were emotions involved.''}). Also, participants described the LLM-integrated interaction as more emotionally engaging (e.g., \textit{``The interaction was somewhat emotional and eye-opening, reinforcing the importance of empathy, active listening, and proper intervention\ldots{}''}). This aligns with LLM’s capability to generate emotional dialogues \cite{mishra2023real, llanes2024developing, liu2024speak}. The results of the thematic analysis of participants’ qualitative feedback revealed additional insights into our findings. This reinforces the conclusion that the LLM-integrated system provided a more engaging and believable experience for participants.

On the other hand, the short-ASAQ percentile rating falls within the lower 25\% for the LLM-integrated system and 5\% for the rule-based system (Figure \ref{fig:asa_comparison}). In contrast to the 29 agents in the ASAQ Representative Set 2024, which are primarily assistant agents, our agent is designed as a training simulation. Therefore, lower ratings in some cognitive constructs, such as performance, attentiveness, and appearance suitability, do not necessarily indicate that the simulation is ineffective. Rather, they may reflect how participants experienced a virtual child in distress, who sometimes "doesn’t know" what to do and needs guidance during the conversation—something a counsellor is expected to provide, not a statement about the system. Response time may have affected attentiveness and performance, as 16 participants noted delays with both agents. However, this is not necessarily a limitation, as real children may also take time to respond. Nevertheless, for training purposes, faster replies might make the experience more engaging. Regarding appearance suitability, one possible explanation is the lack of embodiment, which provides fewer cues during interaction and results in a lack of visible appearance. To gain deeper insights beyond the percentile comparison, the virtual child simulation could be compared more specifically to other chat agents used in social situation simulations of non-experts.

Regarding the experiment's limitations, the LLM agent had noticeable response delays due to server constraints. Although we matched the response timing in the rule-based agent, participants still reported both systems as slow. For the same reason, we used a lightweight model (LLaMA 3.2) instead of the intended LLaMA 3. Initial tests with the larger model suggest improved comprehension and response quality, which could further enhance system performance. Additionally, interactions in our experiment were limited to single 15-minute sessions. Extending session durations or incorporating repeated interactions might reveal more complex dynamics or improved learning outcomes \cite{kang2016spaced}. Future research could also isolate specific components (e.g., NLU or the bypass mechanism) to assess their individual contributions to learning outcomes and trainee experience. This could provide deeper insights into how the system is perceived and its overall effectiveness.

\section{Overall Discussion and Conclusions}

This paper presents a hybrid conversational agent that integrates LLM capabilities into a BDI-based rule system designed for child helpline training. First, we found results that support the non-inferiority of LLM-generated dialogues compared to those generated by humans. Then, through both qualitative and quantitative evaluation, we showed that integrating LLM components for NLU, NLG, and bypass improved believability and trainee attitudes, though the support was weaker for engagement.

Several directions for future work could be considered. One direction is to generalise this approach to other systems that train trainees to change individual beliefs and states in conversations, e.g., training social workers to persuade a virtual agent to get vaccinated, or to de-escalate an aggressive agent. This setup reduces the effort required to model conversation content and responses, particularly when crafting training data or generating response variations. However, it still requires designing the rule-based BDI model to capture the pedagogical constraints and structure of conversations, as these are closely tied to the simulation’s learning objectives. Furthermore, LLMs could be used to generate belief, desire, and intention contents for a scenario \cite{antunes2023prompting}.

LLMs controlling or directly influencing the BDI model can be challenging. LLMs are not explicitly trained to convert inputs into structured cognitive states within a numerical or symbolic framework. As a result, biases inherent in LLMs pose challenges when simulating a BDI conversational model. These include formatting and token biases \cite{jiang2024peek,long2024llms}, as well as issues from unconventional prompts (e.g., prompts including negation \cite{shu2023you}). Therefore, to simulate virtual agents using a BDI framework, LLMs should be paired with a structured BDI system for the LLM to interact with, rather than replace. For example, Retrieval-Augmented Generation \cite{lewis2020retrieval} could be used to dynamically handle BDI-related content, e.g., retrieving relevant beliefs, desires, and rules during interaction.

Although we focused on integrating LLMs into the agent simulator component, the ARTES architecture \cite{alowayyed2024} highlights other potential integration points. For example, LLMs could be used to generate feedback based on trainees’ interactions, enabling deeper skill acquisition \cite{brugge2024large, yao2024enhancing, cook2025virtual}. In our case, the BDI system could support feedback generation by supplying the LLM with structured inputs, such as the child’s internal state, conversational logs, and learning materials. Furthermore, an LLM could simulate a senior counsellor, offering additional information and answering trainees’ queries during interactions.

To conclude, the hybrid approach maintained pedagogical control over the conversational structure, yet offered a better balance between realism and control. This balance opens up new opportunities for scalable, realistic training in sensitive interaction domains, while still aligning with specific learning objectives.

\begin{acks}
This work builds on Adarsh Denga’s Master’s thesis \cite{denga2025combining}. We thank Maya Elasmar, the ProtectMe Consortium, and De Kindertelefoon for their support. The first author’s research is funded by King Saud University and the Saudi Arabian Cultural Mission (SACM).

\end{acks}

\bibliographystyle{ACM-Reference-Format}
\bibliography{sample-base}


\end{document}